\documentclass[twocolumn,showpacs,preprintnumbers,amsmath,amssymb]{revtex4}

\usepackage{graphicx}
\usepackage{dcolumn}
\usepackage{bm}
\usepackage{natbib}

\newcommand{\ket}[1]{\ensuremath{\left|#1\right\rangle}}

\begin{document}

\preprint{APS/123-QED}

\title{Experimental characterization of Gaussian quantum discord generated by four-wave mixing}

\author{Ulrich Vogl}
\author{Ryan T. Glasser}
\author{Quentin Glorieux}
\author{Jeremy B. Clark}
\author{Neil Corzo-Trejo}
\author{Paul D. Lett}
\affiliation{
{Quantum Measurement Division, National Institute of Standards and Technology and}\\
Joint Quantum Institute, NIST \& the University of Maryland, Gaithersburg, Maryland 20899 USA\\
}

\date{\today}

\begin{abstract}
We experimentally determine the quantum discord present in two-mode squeezed vacuum generated by a four-wave mixing process in hot rubidium vapor. The frequency spectra of the discord, as well as the quantum and classical mutual information are also measured. In addition, the effects of symmetric attenuation introduced into both modes of the squeezed vacuum on the discord, the quantum mutual information and the classical correlations are examined experimentally.  Finally, we show that due to the multi-spatial-mode nature of the four-wave mixing process, the quantum discord may exhibit sub- or superadditivity depending on which spatial channels are selected.
\end{abstract}


\maketitle

Quantum communication and computation schemes have been thought to rely on entanglement to offer improvements over classical systems.  Entanglement as a necessary resource can be limiting for experimental implementations of such schemes, since entanglement is easily degraded.  More recently, a different measure of the degree of quantumness of correlations that a bipartite system contains has been proposed - the quantum discord, which allows an appropriate characterization of the quantumness of mixed states \cite{2002Ollivier,2001Henderson,2010Ferraro,Brukner2010,2010Datta,PhysRevX,2011bruss}.  This quantity is defined as the difference between two classically equivalent expressions for the mutual information, when applied to a quantum system.  It has since been shown that systems that are not entangled, but exhibit nonzero quantum discord, still allow for quantum enhancements in a variety of situations \cite{White2008,2011Datta,2011Dowling}. This suggests that quantum discord is, in some scenarios, a more faithful measure of nonclassicality, as it is sensitive to correlations that are ignored by measures like the inseparabilty or the logarithmic negativity \cite{2010Datta,PhysRevX}.  This may allow more robust systems for quantum information processing to be recognized. 

Despite the rapidly growing body of theoretical works concerning quantum discord, few experiments have investigated the topic \cite{2011Yuri,2011Auc,2011Laflamme,Andersen2012,pingkoy2012,grangier2012}.  This may be partly due to the fact that analytic procedures to determine discord in experimental systems involving mixed states are not yet as established as for pure states.
Here we measure the amount of quantum discord, as well as quantum mutual information and classical mutual information, that are present in twin beams created by the vacuum-seeded four-wave mixing (4WM) process in hot Rb vapor. This is based on experimental reconstruction of the covariance matrix of the system.  Additionally, we demonstrate the relative robustness of discord as a nonclassicality measure, as evidenced by the fact that even for vanishing squeezing, non-zero quantum discord still exists.  Finally, we show that the multi-spatial-mode nature of 4WM results in dramatically different behavior of the discord depending on which pairs of spatial sub-channels of the system are used.

We begin by summarizing the concept of mutual information as applied to both classical and quantum systems, and introduce the concept of quantum discord.  Classical information theory results in two equivalent expressions for the mutual information present in a bipartite system with two random variables $A$ and $B$, which take on the values $a$ and $b$, with the probabilities $p_A(a)$ and $p_B(b)$, respectively. The joint probability $p_{AB}(a,b)$ of the system characterizes the correlations between the systems $A$, $B$. The two alternative formulations for the mutual information are \cite{2002Ollivier}:
\begin{align}
\label{1}
I_M(A:B)=H(A)+H(B)-H(A,B)~~~~~~ \text{and} \nonumber \\
J(A:B)=H(B)-H(B|A).~~~~~~~~~~~~~~~~~~~~~~~
\end{align}
Here $H(X)=-\sum p(x) \log_2 p(x)$ is the marginal entropy of the probability distribution for a given result $X$, $H(A,B)$ is the joint marginal entropy, and $H(B|A)$ is the conditional entropy, describing  the average entropies of $B$ conditioned on the alternative outcomes of $A$ following Bayes's rule.
Equation (1) determines the amount of information one can gain about subsystem $A$ upon measurement of subsystem $B$.
Quantum discord results from the fact that that the two classical formulations of mutual information generally do not coincide for a state represented by the density matrix $\rho_{AB}$. In the quantum case, the marginal entropy is replaced by the von Neumann entropy $S(\rho)=-\text{Tr}(\rho \log_2 \rho)$ and yields
\begin{align}
\label{IM}
I_M(\rho_{AB})=S(\rho_A)+S(\rho_B)-S(\rho_{AB}).
\end{align}
The expression for $J(A:B)$ contains the conditional entropy $H(B|A)$, which, when applied to the quantum case results in a term that requires minimization over all possible measurements on subsystem $A$ (due to the ambiguity in defining the state of subsystem $B$ before measuring subsystem $A$).  In general, for the quantum case $I_M\,\neq\,J$, as here measurements on only one subsystem do not give full knowledge about all of the existing correlations.  The difference is the discord $\mathcal{D}=I_M-J$, given by
\begin{align}
\mathcal{D}(A;B)=H(\rho_A)-H(\rho_B)+\underset{A_k}{\text{inf}}\sum_k p_{A_k}H(\rho_{B|k}),
\end{align}
where the infimum describes the optimization over all possible measurements performed on system $A$.  This term makes it an open problem to find an applicable expression under many experimental situations.

To calculate the quantum discord present in a given system assumptions about the state of the system must be made.  We now concentrate on the special case of discord for two-mode Gaussian states, for which an analytic solution for the general case has been given in \cite{2010Datta} and, specifically for symmetric squeezed thermal states, in \cite{2010Paris}. The formalism requires knowledge of the covariance matrix of the bipartite state. In the following we show explicitly how to derive the quantum discord from experimentally obtainable data, based on the theoretical description of Datta et al. \cite{2010Datta}.

A two-mode Gaussian state $\rho_{AB}$ can be described by its covariance matrix $\gamma_{AB}$, which may be transformed into the standard form \cite{2000Duan,Lloyd} with diagonal sub-blocks
\begin{align}
\label{covmat}
\gamma_{AB} =
\left(
\begin{array}{*{4}{c}}
n & 0 & c_1 & 0 \\
0 & n & 0 & -c_2 \\
c_1 & 0 & m & 0 \\
0 & -c_2 & 0 & m\\
\end{array}
\right),
\end{align}
 with the symplectic invariants $I_1=n^2$, $I_2=m^2$, $I_3=c_1c_2$ and $I_4=(nm-c_1^2)(nm-c_2^2)$
 and symplectic eigenvalues
 \begin{align}
 d_{\pm}=\sqrt{\frac{\Delta\pm \sqrt{\Delta^2-4I_4}}{2}},
 \end{align}
where $\Delta=I_1+I_2+2I_3$.

The values $n$, $m$, $c_1$ and $c_2$ are experimentally accessible via homodyne and joint-homodyne measurements, as discussed in the following.
This allows one to express the quantum mutual information, the classical correlations and the Gaussian quantum discord as a function of the symplectic invariants \cite{2010Datta,2010Paris}:
\begin{align}
\label{IM2}
&I_M=h(\sqrt{I_1})+h(\sqrt{I_2})-h(d_+)-h(d_-),\\
\label{J}
&J=h(\sqrt{I_1})-h(\sqrt{E^{min}_{A|B}}),\\
\label{D}
&\mathcal{D}_{A;B}(\gamma_{AB})=h(\sqrt{I_2})-h(d_-)-h(d_+)+h(\sqrt{E^{min}_{A|B}})
\end{align}
with
\begin{align}
\label{infdet}
E^{\min}_{A|B} =
\left\{ \begin{array}{ll}
 \frac{{2 I_3^2 - \left(I_1-4I_4\right) \left(I_2-\frac{1}{4}\right)
 + 2 |I_3| \sqrt{I_3^2 -
 \left(I_1-4I_4\right)\left(I_2-\frac{1}{4}\right)}}}
 {4\left(I_2-\frac{1}{4}\right)^2} \\
\hspace{3cm} \quad\hbox{if }\;
\frac{\left (I_1I_2 - I_4  \right)^2}
{\left(I_1 + 4I_4 \right) \left(I_2 + \frac{1}{4} \right) I_3^2} \le 1 \,, \\[3ex]
 \frac{{I_1 I_2 - I_3^2 + I_4-
 \sqrt{I_3^4+\left(I_1 I_2 - I_4 \right)^2 - 2 I_3^2
 \left(I_1 I_2+I_4\right)}}}{{2 I_2}}\\
 \hspace{5cm} \quad\hbox{otherwise,}
\end{array} \right.
\end{align}
where $h(x)=(x+\frac{1}{2})\log_2(x+\frac{1}{2})-(x-\frac{1}{2})\log_2(x-\frac{1}{2})$.
$\mathcal{D}>1$ is a sufficient condition for demonstrating the entanglement of Gaussian states and $\mathcal{D}=0$ is reached for classical states. A condition arising from the uncertainty relation requires $d_-\geq 1/2$ in order for a state to be physical \cite{2010Datta}.

\begin{figure}
\includegraphics[width=8.5cm]{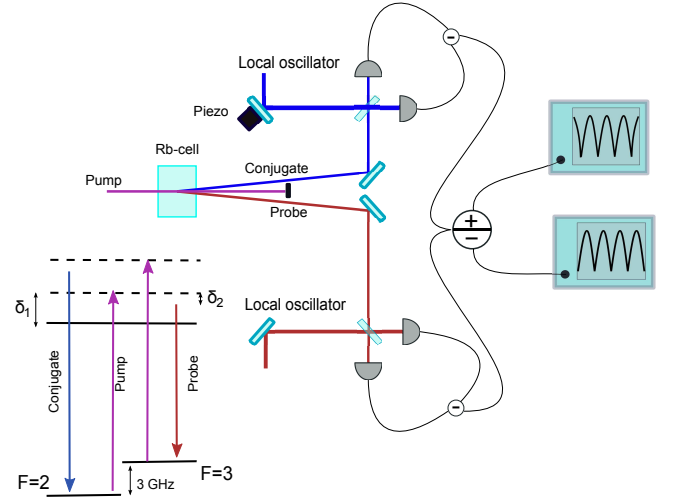}
\caption{\label{fig:epsart} Experimental setup for determining the elements of the covariance matrix. Two-mode squeezed vacuum is generated by a 4WM process. Probe and conjugate beams are generated in a cone around the pump beam with an angle $\approx 1^o$. Each beam is separately overlapped with a strong local oscillator beam on a 50/50 beam splitter, where both outputs are fed into a photodetector.  The relative phase of the local oscillators is scanned by a piezo mirror in one path. The sum and difference of the generated photocurrent signals are fed into electronic spectrum analyzers. $\delta_1$ and  $\delta_2$ are the one- and two-photon detunings, respectively. }
\end{figure}
\begin{figure}[t]
\includegraphics[width=8.5cm]{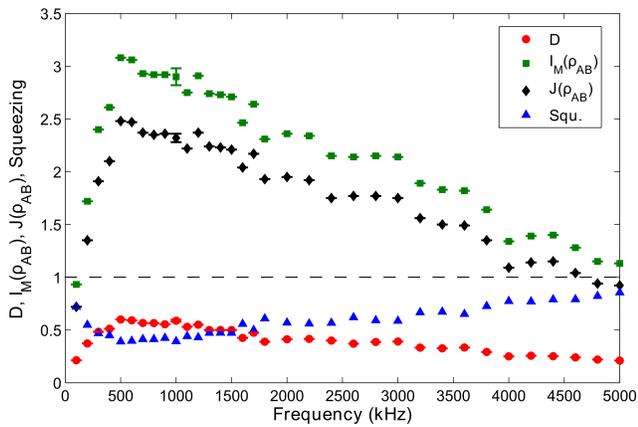}
\caption{\label{fig:epsart} Frequency spectrum for the extracted discord (red circles), quantum mutual information $I_{M}(\rho_{AB})$ (green squares), classical correlations $J(\rho_{AB})$ (black diamonds) and squeezing (blue triangles), respectively (the shot noise limit is indicated with a dashed line).  These measurements, taken at different RF center-frequencies shown, serve to vary the off-diagonal covariance matrix elements (that depend on the amount of squeezing), while effectively keeping the on-diagonal elements constant (these depend only on the excess noise in each single mode).}
\end{figure}

The formulation of discord in Eqs.\,(8,9)
allows for direct experimental access via homodyne and dual-homodyne detection \cite{2005Wenger,2004Bowen,Laurat2005,Paris2010}.  Two-mode squeezed vacuum is generated in multiple spatial modes by a 4WM process near the $^{85}$Rb D1 line $\ket{5S_{1/2},F=2}\rightarrow\ket{5P_{1/2}}$, at $\lambda\approx$ 795\,nm as described in \cite{2009Marino} and indicated in Fig.\,1.; probe and conjugate modes are generated at $\pm$3\,GHz relative to the pump. The one-photon detuning is $\delta_1 \approx$800\,MHz and the two-photon detuning $\delta_2$ is effectively 0\,MHz, as the detected field is determined by the local oscillator detuning. Bright local oscillator beams are created by seeding a second 4WM process that has otherwise the same experimental parameters as the first 4WM process, as in \cite{2009Marino}.

The case examined here involves a symmetric two-mode squeezed state which allows for a simple treatment, resulting in direct experimental access to the nonzero matrix elements in Eq.\,\eqref{covmat}.  A single homodyne detection of each of the modes $A$ and $B$ gives the excess noise fluctuations of the individual modes
\begin{equation}
 \Delta^2X_A ~~\text{and}~~ \Delta^2X_B
\end{equation}
  which, when scaled by the shot-noise limit (SNL), results in the matrix elements $n$ and $m$, respectively.  The off-diagonal matrix elements $c_{1}$ and $c_{2}$ are extracted from measurements of the fluctuations of the joint-amplitude and phase quadratures (difference and sum, respectively):
\begin{align}
X_-=(X_A-X_B)/\sqrt{2},
~~~~~Y_+=(Y_A+Y_B)/\sqrt{2}.~~ 
\end{align}
Here $X_A$, $X_B$ and  $Y_A$, $Y_B$ are the amplitude and phase quadratures of states $A$ and $B$, respectively.  Experimentally, this corresponds  to two separate measurements of the sum and difference signals of a joint-homodyne detection of both modes, while scanning the path length of one of the local oscillators relative to the other, as indicated in Fig.\,1.  These photocurrents give direct access to the variances of the two joint quadratures $\Delta^2X_-$ and $\Delta^2Y_+$ when the local oscillator is scanned and the trace reaches a minimum.   We can then directly determine the off-diagonal entries of the covariance matrix
\begin{equation}
c_1=\Delta^2X_A-\Delta^2X_- ~~\text{and}~~ c_2=\Delta^2Y_B-\Delta^2Y_+,
\end{equation}
after normalizing each quantity to the SNL.

We experimentally examine the frequency spectrum of the discord in our system by changing the RF center-frequency at which the measurements are made. The joint quadrature squeezing and the excess noise on the individual probe and conjugate beams are expected to exhibit a somewhat different frequency dependence due to the variation of the extra noise from the 4WM process.  The measured spectra of discord,  $I_{M}$, $J$, and squeezing are shown in Fig.\,2. The noise variance (blue triangles), normalized such that 1 corresponds to the SNL and 0 is perfect squeezing, exhibits excess noise at low frequencies due to technical noise. Optimum squeezing is obtained in the range $\approx$ 500\,kHz to 1\,MHz, with decreasing measured squeezing at higher detection frequencies.  (All uncertainties shown are one standard deviation, combined statistical and systematic uncertainties.  The statistical uncertainty is obtained from the variance on the measured shot noise and excess noise on the single beam measurements and the joint quadrature measurements. In Fig.\,2 the uncertainties are shown at only one frequency, but are representative of typical uncertainties.)
Though not shown in the plot, there exists nonzero discord for some distance after squeezing has reached zero.   The frequency spectrum of the squeezing is largely reflected in the spectra of the discord (red dots), the quantum mutual information (green squares) and the classical correlations (black diamonds), showing that each of these measures are equally good for investigating the qualitative variation of bipartite correlations. 

\begin{figure}[t]
\includegraphics[width=8.5cm]{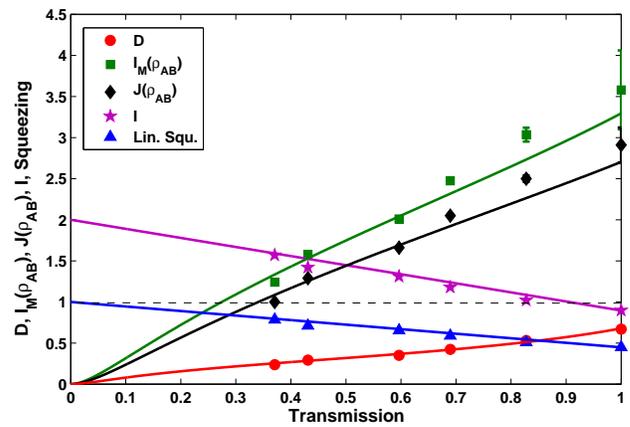}
\caption{\label{fig:epsart} Quantum discord (red circles), inseparability $I$ (purple stars), quantum mutual information $I_{M}(\rho_{AB})$ (green squares), classical correlations $J(\rho_{AB})$ (black diamonds) and squeezing (blue triangles, plotted on a linear scale) versus the (symmetric) transmission of each two-mode squeezed vacuum path when attenuated with neutral density filters.  $I$ at each point is $<2$, confirming that the state is inseparable.  The theory curves are calculated by using the measured initial squeezing and excess noise at a transmission of one. }
\end{figure}

The quantum discord captures information about all of the elements of the covariance matrix.  Another measure, the inseparability criterion, defined as $I=\langle\Delta {X}_{-}^{2}\rangle+\langle\Delta {Y}_{+}^{2}\rangle$, only contains information about the off-diagonal elements.
We now examine how the discord, $I_{M}$, $J$, and $I$ scale with the symmetric attenuation of the two-mode squeezed vacuum state. Before the double-homodyne measurement, we symmetrically vary the amount of attenuation in both modes of the two-mode squeezed vacuum state with a neutral density filter.  This serves to change all of the elements of the derived covariance matrix in a controllable, symmetric manner.  In Fig.\,3 we plot the quantum discord (red dots), the inseparability (purple stars), $I_{M}$ and $J$ versus the transmission of the state, where we attenuate both beams identically with neutral density filters.
        The state is inseparable for all values of transmission measured, according to the criterion of $I\,<\,2$.  The discord, $I_{M}$, and $J$ are all monotonically increasing functions of transmission, and diverge from one another as the transmission is increased. The theoretical curves in Fig.\,3 are calculated for varying attenuation (with the standard beam splitter model for losses \cite{Beamsplitter}) at a transmission of one, with no free parameters. While the inseparability and the squeezing (blue triangles) show linear behavior versus attenuation, the discord curve is slightly nonlinear, indicating the influence of the on-diagonal elements of Eq.\,(4).

\begin{figure}[t]
\includegraphics[width=8.5cm]{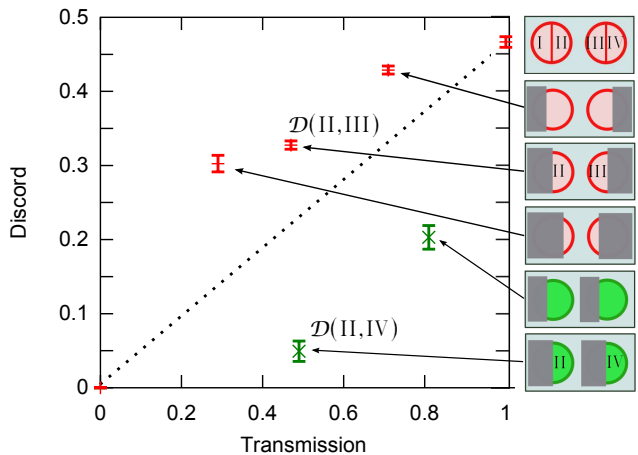}
\caption{\label{fig:epsart} Discord as a function of transmission of the local oscillator beams. The local oscillator beams were clipped symmetrically (red crosses) and antisymmetrically (green stars) as sketched on the right. The points for $\approx$ 50\% clipping are indicated, and the insets have the subregions for 50\% clipping labelled with Roman numerals. }
\end{figure}

The present 4WM system can exhibit multi-spatial-mode twin-beam correlations between the probe and conjugate beams \cite{2008Boyer}. We now investigate the discord and quantum mutual information that is present between various bimodal sub-channels of the two-mode squeezed vacuum. Our aim is to investigate in this context the subadditivity of information measures and channel capacity \cite{Cerf1998,Holevo2012}. 
Subadditivity of a channel is defined here as the situation when the discord of the total bipartite channel is smaller than the sum of the discord between two bipartite sub-channels. Superadditivity is then the converse.
Let us define four channels, made up by combinations of one of the regions I and II as the left and right halves of the probe mode, and one of the regions III and IV as the left and right halves of the conjugate mode (as shown in Fig.\,4).
   The correlations between spatial regions in the probe mode and the conjugate mode are expected to be centro-symmetric with respect to the pump beam, as indicated in Fig.\,1, which allows us to differentiate between mostly correlated and largely uncorrelated subregions.

  We measure the quantum discord, $\mathcal{D}(II;IV)$, present between modes II and IV by cutting the local oscillator beams with a knife-edge, and only allowing these modes to be fed into the dual-homodyne detection system.  This serves to project out only modes II and IV.  As seen in Fig.\,4, the quantum discord present between these two channels, $\mathcal{D}(II;IV)$, is much less than half of the total discord present between the entire, uncut probe and conjugate modes, $\mathcal{D}(I+II;III+IV)$.  In this case, the discord is superadditive, or in other words, we have chosen subregions of the beams that are largely uncorrelated. (The same result is obtained with $\mathcal{D}(I;III)$, confirming the spatial symmetry when clipping the beams in this direction). In contrast, the discord between modes II and III, $\mathcal{D}(II;III)$, is significantly more than half of the total discord between the total, uncut modes.  In this scenario, the discord behaves subadditively.  This subadditive property \cite{Cerf1998} of the discord may allow for more effectively exploiting the channel capacity of a total system if it is correctly divided into various bimodal sub-channels.  This shows that the discord varies dramatically in multi-spatial-mode squeezed vacuum systems, depending on which spatial channels are utilized. Here, when symmetric channels are used, the discord is subadditive with respect to the unaltered system.  We also find that when asymmetric channels are chosen, the amount of discord quickly decreases, resulting in superadditive behavior.

Existing theoretical work concerning quantum discord indicates that discord can in principal be employed in various quantum computation and quantum communication schemes to provide enhancements beyond those allowed classically.  In particular, many schemes require optical delay and storage of quantum correlations that are always accompanied by some loss and decoherence.  The relative robustness of quantum discord with respect to decoherence effects provides motivation to more closely investigate these systems experimentally.  Specifically, the present scheme for the determination of discord is applicable if one is interested in the evolution of quantum correlations in systems that involve stored light \cite{quentin2012} and fast light \cite{2012PRL,fastinfo}, where squeezing and entanglement may be easily degraded. Discord as a measure of quantum correlations in mixed bipartite states may offer a means to experimentally test the non-broadcasting theorem for the class of quantum-correlated bipartite states as a simple example of mixed states \cite{nobroadcast,nobroadcast2} under fast-light conditions, and how classical and discordant states can be distinguished in this regime.   Examining the amount of discord, rather than squeezing or entanglement, present under these exotic conditions should allow for a more easily experimentally obtainable measure of any ``quantumness" that is present in the system.

\begin{acknowledgments}
This work was supported by the Air Force Office of
Scientific Research. Ulrich Vogl would like to
thank the Alexander von Humboldt Foundation. This research was performed while
Ryan Glasser held a National Research Council Research
Associateship Award at NIST.
\end{acknowledgments}


\end{document}